# Parallelize Bubble Sort Algorithm Using OpenMP

Zaid Abdi Alkareem Alyasseri
IT-RDC- University of Kufa
Iraq

Kadhim Al-Attar, Mazin Nasser, ISMAIL
CS School- USM
Malaysia

E-mail: Zaid.alyasseri@uokufa.edu.iq

*Abstract— Sorting has been a profound area for the algorithmic researchers and many resources are invested to suggest more works for sorting algorithms. For this purpose, many existing sorting algorithms were observed in terms of the efficiency of the algorithmic complexity. In this paper we implemented the bubble sort algorithm using multithreading (OpenMP). The proposed work tested on two standard datasets (text file) with different size . The main idea of the proposed algorithm is distributing the elements of the input datasets into many additional temporary sub-arrays according to a number of characters in each word. The sizes of each of these sub-arrays are decided depending on a number of elements with the same number of characters in the input array. We implemented OpenMP using Intel core i7-3610QM ,(8 CPUs),using two approaches (vectors of string and array 3D) . Finally, we get the data structure effects on the performance of the algorithm for that we choice the second approach.*

**Keywords-** *Bubble sort, OpemMP, sorting algorithms, parallel computing, Parallelize Bubble algorithm.*

## I. INTRODUCTION

Sorting is one of the most common operations perform with a computer. Basically, it is a permutation function which operates on elements [1]. In computer science sorting algorithm is an algorithm that arranges the elements of a list in a certain order. Sorting algorithms are taught in some fields such as Computer Science and Mathematics. There are many sorting algorithms used in the field of computer science such as Bubble, Insertion, Selection, Quick..etc. They differ in their functionality, performance, applications, and resource usage[2]. In this paper we will focus on the bubble sort algorithm. Bubble sort is the oldest, the simplest and the slowest sorting algorithm in use having a complexity level of $O(n^2)$. Bubble sort works by comparing each item in the list with the item next to it and swapping them if required. The algorithm repeats this process until to make passes all the way through the list without swapping any items. Such a situation means that all the items are in the correct order. By this way the larger values move to the end of the list while smaller values remain towards the beginning of the list. It is also used in order to sort the array such like the larger values comes before the smaller values [1]. In other words, all items are in the correct order. The algorithm's name, bubble sort, comes from a natural water phenomenon where the larger items sink to the end of the list whereas smaller values "bubble" up to the top of the data set [2]. Bubble sort is simple to program, but it is worse than selection sort for a jumbled array. It will require many more component exchanges, and is just good for a pretty well ordered array. More importantly bubble sort is usually the easiest one to write correctly [4].

## II. PROPOSED METHOD

The main idea of the proposed algorithm is distributing the elements of the input datasets into many additional temporary sub-arrays according to a number of characters in each word. The sizes of each of these sub-arrays are decided depending on a number of elements with the same number of characters in the input array. There are two types of text file dataset have been provided in this paper (HAMLET, PRINCE OF DENMARK by William Shakespeare), which are different in size and length. The first dataset is equal (190 KB) and the second one is equal (1.38 MB) taken from (http://www.booksshouldbefree.com/).

**Pre-processing**

We sort the datasets using the bubble sort algorithm in three phases. In the first phase, we are removing / ignoring the special characters from the text file. In the second phase, we convert the text file to array of list (vectors of string) based on the length of characters, all shorter words come be for longer words. In the third phase, we sort each vector of string by arranging in the alphabetic order using the bubble sort algorithm. Table 1 shows the time of pre-processing phase. The sequential performance has been implemented on ASUS A55V , Windows 7 Home Premium 64-bit with Intel core i7-3610QM ,(8 CPUs), 2.30GH and 8 GB of RAM.

Table 1 Execution Time of Pre-processing phase

| Function | Data Set 1 size of (190KB) | Data Set2 size of (1.38MB) |
|---|---|---|
| Preprocessing to remove the special characters from a text file | 0.265 second | 1.154 second |
| Preprocessing on file by using bubble sort based on alphabetic order | 274.528 second | 274.528 second |
| Preprocessing on file by using bubble sort based on the length of the word | 230.215second | 230.215second |
| Bubble sort on array of vector each vector has the same length of the word | 42.589 second | 1620.44 second |

## III. BUBBLE SORT AS A SEQUENTIAL

The sequential version of the bubble sort algorithm is considered to be the most inefficient sorting method in common usage. In the sequential code we implement two approaches which are difference in the data structure.
1)**Approach one** : Implemented based on vectors of strings
The procedure of first approach: 1.Loading the data from the text file and store in the vector. 2.Removing/Ignoring the special characters ( e.g. ", . ? ! etc.) from the text file. This is the preprocessing stage as we mention above. 3.Creating an array of vector based on the longest word in the text file. 4.Appling the bubble sorting for each word in the vector, the bubble sort is based on ASSCII code for each letter in the text file.

2)**Approach Second** : implemented based on array char 3D
The procedure is like the first one but different in the representation of data inside the memory (the difference in the data structure). The main important step here is how to create three dimensional array , the next code shows how to do that. Finally, We have to do the bubble sort is based on the new method (3D array of char).

The performance of the sequential code for two datasets which has been tested for 10 times to get the average is shown in table 2 and 3.

**Result of the First Approach Based on Dataset 1 &** 2
Table 2   Shows datasets have been tested 10 times to get an average based on the first approach.

| Test / Sec | Dataset1 (190 KB ) | Dataset2 (1.32 MB ) |
| --- | --- | --- |
| T1 | 44.239 | 1694.51 |
| T2 | 44.013 | 1697.98 |
| T3 | 44.239 | 1702.21 |
| T4 | 44.503 | 1696.88 |
| T5 | 44.23 | 1698.3 |
| T6 | 44.746 | 1692.79 |
| T7 | 44.433 | 1687 |
| T8 | 44.36 | 1686.96 |
| T9 | 44.267 | 1674.82 |
| T10 | 44.704 | 1630.32 |
| Ave | 44.373 | 1686.177 |

**Result of the Second Approach Based on Dataset 1 &2**
Table 3   Shows datasets have been tested 10 times to get an average based on the second approach.

| Test / Sec | Dataset1 (190 KB ) | Dataset2 (1.32 MB ) |
| --- | --- | --- |
| T1 | 6.695 | 188.185 |
| T2 | 6.624 | 188.194 |
| T3 | 6.615 | 188.247 |
| T4 | 6.694 | 188.169 |
| T5 | 6.531 | 188.348 |
| T6 | 6.654 | 188.289 |
| T7 | 6.654 | 188.154 |
| T8 | 6.614 | 188.512 |
| T9 | 6.646 | 188.343 |
| T10 | 6.672 | 188.181 |
| Ave | 6.639 | 188.262 |

**BUBBLE SORT AS PARALLEL**
One of the fundamental problems of computer science is ordering a list of items. There are a lot of solutions for this problem, known as sorting algorithms. Some sorting algorithms are simple and intuitive, such as the bubble sort, but others, like quick sort, are extremely complicated but produce lightening-fast results. The sequential version of the bubble sort algorithm is considered to be the most inefficient sorting method in common usage. In this paper we want to prove that how the parallel bubble sort algorithm is used to sort the text file parallel and will show that it may or may not better than the sequential sorting algorithm. The old complexity of the bubble sort algorithm was $O(n^2)$, but now we are using the complexity for bubble sort algorithm n(n-1)/2. Algorithm 1 ( in chapter 1) shows the code for the bubble sort algorithm. As usually in parallelism we can decompose the data or functions or sometimes both, but after we are studying the bubble sort we reached to the point, that it is more suitable for data decomposition. So we suggest during the implementation of OpenMP is to assign each vector to individual process.

**BUBBLE SORT IN OPENMP**
OpenMP is a widely adopted shared memory parallel programming interface providing high level programming constructs that enable the user to easily expose an application's task and loop level parallelism in an incremental fashion. The range of OpenMP applicability was significantly extended recently by the addition of explicit tasking features. The OpenMP is the dominant programming model for heterogeneous systems and adopted by Intel, Clear Speed, PGI and CAPS SA. The idea behind OpenMP is that the user specifies the parallelization strategy for a program at a high level by providing the program code [9]. OpenMP is the implementation of multithreading, a parallel execution scheme where the master thread assigns a specific number of threads to the slave threads and a task is divided between them. OpenMP is basically designed for shared memory multiprocessors, using the SPMD model (Single Program, Multiple Data Stream). All the processors are able to directly access all the memory in the machine, through a logical direct connection. Programs will be executed on one or more processors that share some or all of the available memory. The program is typically executed by multiple independent threads that share data, but may also have some additional private memory zones [10]. OpenMP provides a straightforward interface to write software that can be used for multiple core computers. Using OpenMP the programmer can write code that will be able to use all cores of a multicore computer, and will be run faster if the number of cores are increased. In this section we will implement our sequential bubble sort as parallel by using an OpenMP model in visual C++ language, in the early stage of the program writing we should adapt our environment to be able to understand OpenMP statements. It is not worthwhile just to write #include <omp.h> header in our program but also required to include some configurations to C++ environment. After the creation of our project, we update the C++ environment to familiar with OpenMP as follows: From the Tools menu we select project properties and below the windows appear by which we choose/set configuration properties for C/C++ language, after that we change the "OpenMP Support field" to yes value. Now if we return back to our sequential program the first question can the program inherits the parallelism? After the evaluation of the program we identify that it can be worked as a parallel. The second question which parts of it can be worked as parallel. It depends on understanding the problem, so we found that hot spots occur in the loop statements as more perfectly because we work on our problem as 3D matrix, so we will assign each vector to be implemented in one processor, as a result the vector of 1,2,3.... length will be executed separately at the same time.

```
void bubbleSortAsciiFor3DA(char *a, unsigned int *l,int x, int y, int z) {
    cout << "please enter your threads  no. :- ";
    int thread;
    cin >> thread;
    #pragma omp parallel num_threads (thread)   //  in this part we tell the OpenMP the number of processors
    {
    #pragma omp parallel for   // make the parallel in for statements
    for (int i= 1; i<x;i++)
        {
        bubbleSortAscii3D(a,x,y,z,i,*(l+i));//sorting each vector that contain the same length in the array of vectors
        }
} }
```

Parallel loop: executing each iteration concurrently is the same as executing each iteration sequentially. No loop carries dependencies (an iteration does not produce any data that will be consumed by another iteration).

**OPENMP RESULATS**

Serial and Multithread program verification tools For testing our serial and Multithreading (OpenMP) program we used Intel® Inspector XE. The Intel® Inspector XE is a dynamic memory and threading error checking tool for users developing serial and multithreaded applications on Windows* and Linux* operating systems. You can also use the Intel Inspector XE to visualize and manage static analysis results created by Intel® compilers in various suite products. Where the figure 1 shows the result of applies Intel® Inspector XE. Fig 4 shows testing our program and finlly there not any error there

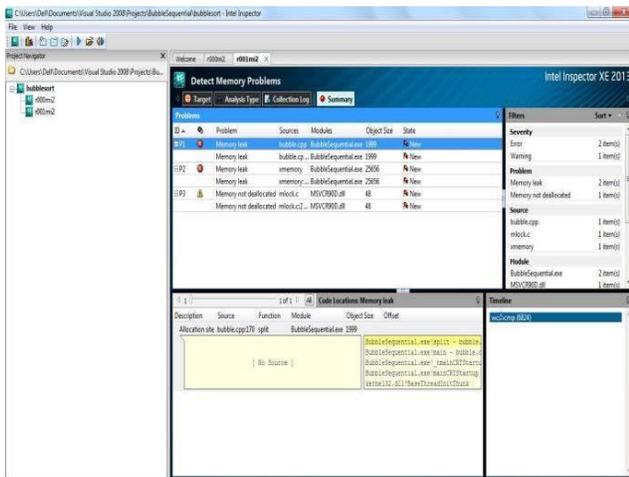

Fig. 1 Show the result of testing our code using Intel ® Inspector XE.
TABLE 4 shows the result of OpenMP for dataset one and two. In this table we use a number of threads (1,2,4,8,10 and 16) and test each thread three times. We compute the speedup and efficiency for Array of Char 3D. For the time we calculate the average time of three time testing and then divide the serial time by parallel time to get the speedup(Ts/Tp). For efficiency we divide speedup on the number of threads (**speed up/No. of threads**).

TABLE 4 BUBBLE SORT WORK AS PARALLEL USING OPENMP

| T. No | Appearance (in Time New Roman or Times) | | | | | |
|---|---|---|---|---|---|---|
| | Dataset 1 Time / Sec | Dataset 2 time / Sec | Speed up1 | Speed up2 | Effici ency1 | Effici ency2 |
| 1 | 6.695 | 188.185 | 1 | 1 | 100% | 100% |
| 2 | 5.103 | 132.662 | 1.311 | 1.418 | 65% | 70% |
| 4 | 4.572 | 84.271 | 1.464 | 2.233 | 36% | 55% |
| 6 | 3.751 | 65.846 | 1.784 | 2.857 | 29% | 47% |
| **8** | **3.167** | **51.046** | **2.113** | **3.686** | 26% | 46% |
| 10 | 3.826 | 52.991 | 1.749 | 3.551 | 17% | 35% |
| 16 | 4.858 | 65.089 | 1.378 | 2.891 | 13% | 18% |

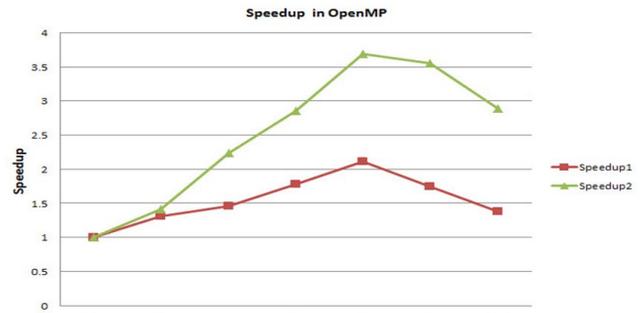

Fig. 2 Show the speedup of dataset1 and 2 using OpenMP

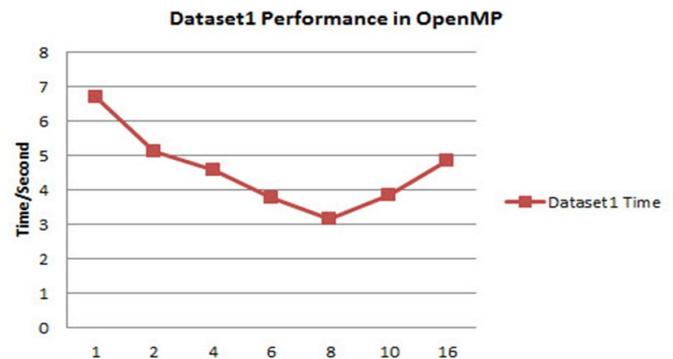

Fig. 3 shows the performance of OpenMP with Dataset 1

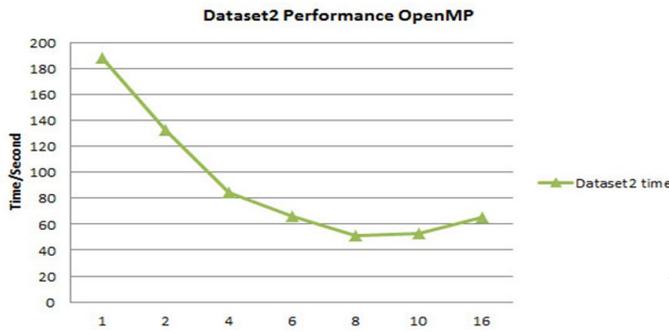
Fig. 4 shows the performance of OpenMP with Dataset 2

### IV. Analysis the result of OpenMP

According to the pervious results the OpenMP shows the best speedup was when used 8 threads. That means, the best speedup occurs when using threads number as equal to the actual cores number (see table 4). In other words, increasing the number of threads up to the actual number of cores do not lead to any advantage really it will be effected on the speedup value. Because the increasing of thread mean more works to dividing the tasks, create threads and destroy it..etc. And what proves this conclusion the Increasing decadence in the value of the efficiency as a result we will have huge idle time.

### V. CONCLUSION

In this paper we implemented the bubble sort algorithm using multithreading (OpenMP). The proposed work tested on two standard datasets (text file) with different size taken from (HAMLET, PRINCE OF DENMARK by William Shakespeare) (http://www.booksshouldbefree.com/). We implemented OpenMP using ASUS A55V , Operating System: Windows 7 Home Premium 64-bit Processor: Intel core i7-3610QM ,(8 CPUs), 2.30GH ,memory: RAM 8 GB. Finally, we get the data structure effects on the performance , where this is clear in sequential code 1 and 2. In OpenMP, increasing the number of threads more than an actual core number it will be effected on the seepd up only. For the future work we will implement bubble sort using massage Passing Interface (MPI) and compiler the resulte with OpenMP approach.

ACKNOWLEDGMENT

The first author would like to thank University of Kufa for supporting this research. Also we would like to thank ***Prof. Rosni Abdullah, Dr. Mohd. Adib Hj. Omar*** and all postgraduate students of Parallel Computing Architectures and Algorithms class from school of computer sciences at University Sains Malaysia for supporting this work.